\documentclass[aps, twocolumn, superscriptaddress, showpacs, longbibliography]{revtex4-1}

\usepackage{amsmath, amsfonts, amssymb, amsthm}
\usepackage{hyperref}
\usepackage{bbm}
\usepackage{physics}
\usepackage{tabularx}
\usepackage{tikz}
\usepackage{usefulnotations}

\newcommand{\dsb}[2]{%
	[[ #1 ]]_{#2}%
}

\begin{document}
	\title{Non-Hermitian Generalization of Rayleigh-Schr\"{o}dinger Perturbation Theory}

	\author{Wei-Ming Chen}
	\affiliation{Department of Physics, National Sun Yat-sen University, Kaohsiung 80424, Taiwan}
	\affiliation{Center for Theoretical and Computational Physics, National Sun Yat-sen University, Kaohsiung 80424, Taiwan}
	\author{Yen-Ting Lin}
	\affiliation{Department of Physics, National Sun Yat-sen University, Kaohsiung 80424, Taiwan}
	\affiliation{Center for Theoretical and Computational Physics, National Sun Yat-sen University, Kaohsiung 80424, Taiwan}
	\author{Chia-Yi Ju}
	\email{chiayiju@mail.nsysu.edu.tw}
	\affiliation{Department of Physics, National Sun Yat-sen University, Kaohsiung 80424, Taiwan}
	\affiliation{Center for Theoretical and Computational Physics, National Sun Yat-sen University, Kaohsiung 80424, Taiwan}

	\begin{abstract}
		While perturbation theories constitute a significant foundation of modern quantum system analysis, extending them from the Hermitian to the non-Hermitian regime remains a non-trivial task.  In this work, we generalize the Rayleigh-Schr\"{o}dinger perturbation theory to the non-Hermitian regime by employing a geometric formalism.  This framework allows us to compute perturbative corrections to eigenstates and eigenvalues of Hamiltonians iteratively to any order. Furthermore, we observe that the recursion equation for the eigenstates resembles the form of the Girard-Newton formulas,  which helps us uncover the general solution to the recursion equation. Moreover, we demonstrate that the perturbation method proposed in this paper reduces to the standard Rayleigh-Schr\"{o}dinger perturbation theory in the Hermitian regime.
	\end{abstract}

	\maketitle
	\newpage

	\section{Introduction}

		It is well-known that perturbation theories are crucial tools to analyze quantum systems.  Not only do they yield good approximations near analytically solvable systems, but they also provide insights into quantum systems~\cite{SHIJIAN2010, Jain2020, Tzeng2021, Tu2023, Lin2024}. It is safe to say that much of our understanding of quantum physics relies on the perturbation theories.

With this in mind, as non-Hermitian systems~\cite{Bender1998, Mostafazadeh2003, Bender2004, Bender2007, Mostafazadeh2010, Brody2013} become increasingly popular and intriguing~\cite{Tu2022, Ju2022, Znojil2023, Znojil2023a, Znojil2023b, Abo2024, Arkhipov2024, Yang2024, Znojil2024, Rakhmanov2024},  perturbation results are expected to shed light from different perspectives.  For instance, many studies aim to generalize the perturbation theories, particularly the Rayleigh-Schr\"{o}dinger perturbation theory,  but they are often case studies or still in the process of laying the groundwork for a formal non-Hermitian framework~\cite{Caliceti1980, Ibanez2014, Znojil2020, Znojil2020a, Znojil2024a}. Thus,  in contrast to the well-established perturbation methods in the Hermitian regime,  a systematic extension into the non-Hermitian regime has yet to be fully developed, and as a result, an organized treatment is desired.  

		Therefore, in this work, we provide a generalization of the Rayleigh-Schr\"{o}dinger perturbation method~\cite{Schroedinger1926} using the geometric formalism of quantum mechanics~\cite{Mostafazadeh2004, Ju2019}.  The generalization is applicable to both Hermitian and non-Hermitian regimes.  In the geometric formalism, the parameters of a quantum system naturally induce extra dimensions in which the states evolve~\cite{Ju2024}. Consequently, perturbations in the states correspond to infinitesimal evolutions in the parameter-induced dimension. This makes the perturbation framework more intuitive and, therefore, easier to tackle.  Additionally,  the framework enables us to derive a concise form of the perturbed eigenstates and eigenvalues using the generalized resemblance to the Girard-Newton formulas.

		 The paper is organized as follows. In Sec.~\ref{secII}, we give a brief review of the geometric formalism of quantum mechanics and the emergent dimensions. In Sec.~\ref{secIII}, the materials mentioned in Sec.~\ref{secII} are applied to the general idea of the perturbation theory. Then, in Sec.~\ref{secIV}, we focus on the standard linear-perturbation scheme and compare it with the one in the Hermitian regime. To make the formalism developed in Sec.~\ref{secIII} more clearly, a non-trivial example is given in Sec.~\ref{secV}. Finally, we conclude the paper with discussions in Sec.~\ref{secVI}.

	\section{Geometric Formalism of Quantum Mechanics}\label{secII}

		In this section, we review the geometric formalism of (non-)Hermitian quantum mechanics~\cite{Ju2019, Ju2022} and its direct consequence, namely, an emergent dimension~\cite{Ju2024}. In this context, the physical system under consideration evolves with respect to the parameters $t$ (time) and $q$ (a physical parameter) as parallel transports.

		To be more specific, the states in the Hilbert space are constrained by the Schrödinger equation, i.e., a parallel transport in the $t$-direction,
		\begin{equation}
			\label{eqH} \nabla_t |\psi(q,t)\rangle = \bigg[ \frac{\partial}{\partial t} + i H(q,t)\bigg]|\psi(q,t)\rangle = 0,
		\end{equation}
		where $\nabla_t$ is the connection or covariant derivative in the $t$-direction.
		An important aspect of this formalism is the Hilbert space equipped with the metric $G(q,t)$, which enables the definition of a more general inner product in the Hilbert space,
		\begin{equation} 
			\langle\psi_1(q,t)| G(q,t)|\psi_2(q, t)\rangle,
		\end{equation}
		where $\bra{\psi_1(q, t) }$ is the Hermitian conjugate of $\ket{\psi_1(q,t)}$. Note that when $G(q, t)$ is the identity, this definition reduces to the familiar form encountered in Hermitian quantum systems.
		 To express this more concisely, we define
		\begin{equation}
			\Ket{\psi(q, t)} = \ket{\psi(q, t)}\quad \text{and} \quad \Bra{\psi(q, t)} = \bra{\psi(q, t)} G(q, t).
		\end{equation}
		We therefore treat the double brackets $\Ket{\psi(q,t)}$ and $\Bra{\psi(q,t)}$ as states and dual state, respectively, in a Hilbert space.
		Moreover, by requiring that the state normalization is independent of the parameters $q$ and $t$, i.e.,
		\begin{equation}
			\label{pind} \frac{\partial}{\partial q}\Braket{\psi(q,t)} {\psi(q,t)} = \frac{\partial}{\partial t}\Braket{\psi(q,t)} {\psi(q,t)} = 0,
		\end{equation}
		we obtain the governing equations on the metric $G(q, t)$ being
		\begin{equation}
			\nabla_t G(q, t) = i G(q, t) H(q, t) - i H^\dagger(q, t) G(q, t),
		\end{equation}
		and
		\begin{equation}
			\nabla_q G(q, t) = i G(q, t) K(q, t) - i K^\dagger(q, t) G(q, t),
		\end{equation}
		and the state evolution along the $q$-direction being
		\begin{equation}
			\label{eqK} \nabla_q \ket{\psi(q,t)} = \left[ \frac{\partial}{\partial q} + i K(q,t)\right] \ket{\psi(q,t)} = 0.
		\end{equation}
		Although $K(q, t)$ is not yet determined at this stage, from its role indicates that it is the evolution generator in the $q$-direction, or the $q$-generator. It is worth mentioning that even without an explicit expression or governing equation, we can show that the $q$-generator $K(q, t)$ is not unique. As an example, one can add a phase to the state, i.e., $|\psi(q,t)\rangle \rightarrow e^{A q} |\psi(q,t)\rangle$, without altering the physics, but it modifies $K(q, t)$ in Eq.~\eqref{eqK} to $K(q, t) + A$. Therefore, the $q$-generator $K(q, t)$ is gauge-dependent (a more precise argument can be found in~\cite{Ju2019, Ju2024}).

		To find the explicit expression for $K(q, t)$, we turn our attention to the local curvature $\mathcal{F} = F_{t,q} dt \wedge dq$, where
		\begin{equation}
			i F_{t,q} \ket{\psi} = [\nabla_t, \nabla_q] \ket{\psi} = 0,
		\end{equation}
		for any quantum state $\ket{\psi(q, t)}$, where the last equality comes from Eqs.~\eqref{eqH} and \eqref{eqK}.  In other words,  the local curvature $\mathcal{F}$ always vanishes.

		As a consequence, from $F_{t,q} = 0$, we obtain the relation between the operators $H$ and $K$, namely
		\begin{equation}
			\label{eqHK} i \frac{\partial}{\partial t} K(q,t) - i \frac{\partial}{\partial q} H(q,t) + [K(q,t),H(q,t)] = 0.
		\end{equation}
		This equation serves as our starting point for performing the perturbation computation.  As we will see later, the perturbation in the Hamiltonian can be encoded in $K(q, t)$.  While $K(q, t)$ is gauge-dependent,  we can choose suitable gauge choices for eigenstates perturbation.  Hence, each term of the Hamiltonian eigenvalue perturbation can be expressed using the operator $K(q, t)$.

	\section{Geometric Perturbation}\label{secIII}

		For simplicity, we focus on the non-degenerate perturbation theory for the eigenstates of a time-independent Hamiltonian. In this case, the time evolution of the eigenstates can be written as
		\begin{equation}
			\label{state}\Ket{\psi_n(q,t)} = e^{ - i h_n(q) t}\Ket{n(q)},
		\end{equation}
		where $\ket{n(q)} = \ket{\psi_n(q, t = 0)}$ is the $n$-th eigenstate of the Hamiltonian, i.e.,
		\begin{equation}
			\label{Hamil} H(q) \Ket{n(q)} = h_n(q) \Ket {n(q)} ,
		\end{equation}
		with $h_n(q)$ representing the $n$-th eigenvalue of the Hamiltonian. To find a $q$-generator $K(q, t)$ that works well with the perturbation, we impose the adiabatic gauge-fixing condition~\cite{Ju2024},
		\begin{equation}
			\label{gfix} \left[\partial_t K(q,t) , H(q) \right] = 0.
		\end{equation}
		This choice ensures that the states evolve adiabatically, such that Eq.~\eqref{Hamil} is preserved. The condition, together with Eq.~\eqref{eqHK}, leads to
		\begin{equation}
			\partial_t^2 K(q,t) = 0,
		\end{equation}
		which allows us to decompose $K(q, t)$ into
		\begin{equation}
			\label{decom} K(q,t) = K_1(q) t + K_0(q).
		\end{equation}
		From the decomposition, it is easy to see the operator $K_1(q)$ is completely fixed by the gauge-fixing condition in Eq.~\eqref{gfix}, while $K_0(q)$ retains some gauge freedom~\cite{Ju2024a} (i.e., $K_0(q)$ can be shifted by $\Delta K$, where $[\Delta K(q), H(q)] = 0$).

		From now on, since the operators $H(q)$, $K_1(q)$, and $K_0(q)$, as well as the states $\Ket{n(q)}$, depend solely on the parameter $q$, we will omit this dependence in our expressions for simplicity. Additionally,  we will omit the $q$- and $t$-dependence in $K(q, t)$ to further simplify the discussion.

		By substituting the decomposition into Eqs.~\eqref{eqHK} and \eqref{gfix},  we obtain in turn the following commutation relations, respectively:
		\begin{align}
			\label{HK0}\left[H,K_0\right]& = i(K_1 - \partial_q H),\\[2mm]
			\label{HK1}\left[H,K_1\right]& = 0.
		\end{align}

		Since $K_1$ and $H$ commute with each other, they share the same eigenstates. To compute the eigenvalue perturbation, it is important to determine the eigenvalue of the corresponding eigenstate when $K_1$ acts on it.

		By applying the parallel transport in the $q$-direction Eq.~\eqref{eqK} to the time-independent Hamiltonian eigenstate Eq.~\eqref{state},
		\begin{align}
		0 & = \nabla_q \big(e^{ - i h_n t}\Ket{n_q}\big)\\
		\notag & = e^{ - i h_n t}\big( - i t \partial_q h_n + \partial_q + i K\big)\Ket{n}\\
		\notag & = e^{ - i h_n t} \Big[ - i t\big( \partial_q h_n - K_1\big) + \big(\partial_q + i K_0\big)\Big]\Ket{n},
		\end{align}
		where, in the last equality, we have used the gauge-fixing decomposition Eq.~\eqref{decom}. Since terms with different powers in $t$ are independent, this implies,
		\begin{align}
			\label{K01}K_0\Ket{n} & = i\partial_q\Ket{n}, \\[2mm]
			\label{K11}K_1\Ket{n} & = \partial_q h_n\Ket{n}.
		\end{align}
		These results are consistent with the previous discussion: $K_0$ is not a gauge-invariant operator, as it generates $q$-translations in eigenstates (which are not physical observables), whereas $K_1$ is a gauge-invariant operator, as it generates $q$-translations in eigenvalues (which are physical observables).

		Next, we proceed to compute the perturbative corrections to the eigenstates and their corresponding eigenvalues. Assume $q$ is small and expand all relevant quantities in Taylor series,
		\begin{align}
			\notag & K_0 = \sum_{j = 0}^{\infty} q^j \p{K_0}{j},~~
			\label{Expn} K_1 = \sum_{j = 0}^{\infty} q^j \p{K_1}{j}\,, ~~H = \sum_{j = 0}^{\infty} q^j \p{H}{j}\,,\\
			& \hspace{1.3cm}h_n = \sum_{j = 0}^{\infty} q^j h_n^{(j)}\,,~~~\Ket{n} = \sum_{j = 0}^{\infty} q^j \Ket{n^{(j)}}.
		\end{align}
		The upper indices in the parentheses denote the orders in the Taylor expansion.

		Given the expansion, Eq.~\eqref{K01} allows us to express a given state correction as a linear combination of $K_0$ acting on other state corrections. More precisely, for order $q^{k - 1}$, Eq.~\eqref{K01} implies,
		\begin{equation}
			\label{GNFstate} \Ket{\p{n}{k}} = - \frac{i}{k}\sum_{j = 1}^{k} \p{K_0}{j - 1}\Ket{\p{n}{k - j}},~~~~~~ k\geq 1.
		\end{equation}

		This re-expression can be performed iteratively, so that any correction to a perturbed state can be expressed as a linear combination of products of $\p{K_0}{j}$'s acting on the unperturbed state $\Ket{\p{n}{0}}$. Interestingly, Eq.~\eqref{GNFstate} closely resembles the form of the Girard-Newton formulas (see Appendix~\ref{App:GNFandBP}).  Specifically,  it can be obtained by replacing $s_j$ with $ \Ket{\p{n}{j}}$ and $p_j$ with $i \p{K_0}{j - 1}$ in Eq.~\eqref{GNF}, respectively. Therefore, just as $s_j$ can be represented by $p_j$'s and $s_0$, $\Ket{\p{n}{k}}$ can also be expressed in terms of $\p{K_0}{j}$'s and $\Ket{\p{n}{0}}$. However, one must be careful with the ordering: unlike the terms in the standard Girard-Newton formulas, $\p{K_0}{j}$'s are generally non-commutative operators, and the states must also be placed after the $\p{K_0}{j}$'s. Therefore, the ordering of $\p{K_0}{j}$'s must be taken into account when solving for the eigenstates from Eq.~\eqref{GNFstate}. As a result, Eq.~\eqref{GNFstate} can be expressed as
		\begin{widetext}
		\begin{align}
			\label{recurstatesol} & \Ket{\p{n}{k}} = \frac{\mathbb{B}_k\Big(  (0!)\cdot (\p{-i\,K_0}{0}),   ( 1)! \cdot (\p{-i\,K_0}{1}), \dots,   (k - 1)!\cdot (\p{-i\,K_0}{k - 1})\Big)}{k!}\Ket{\p{n}{0}},
		\end{align}
		\end{widetext}
		with $\mathbb{B}_k$ being the \textit{dual non-commutative Bell polynomials}, where the recursive definition can be found in Eq.~\eqref{NCBDef} in Appendix~\ref{App:GNFandBP}.
		For clarity, we provide some examples below:
		\begin{align}
			\notag\Ket{\p{n}{1}} & = - i \p{K_0}{0}\Ket{\p{n}{0}},\\[1mm]
			\Ket{\p{n}{2}} & = - \frac{1}{2}\Big[ \left(\p{K_0}{0}\right)^2 + i \p{K_0}{1}\Big] \Ket{\p{n}{0}},\\[1mm]
			\notag\Ket{\p{n}{3}} & = \frac{i}{6}\Big[\left(\p{K_0}{0}\right)^3 + i \p{K_0}{0} \p{K_0}{1}\\
			\notag &\hspace{1.74cm} \quad+ 2 i \p{K_0}{1} \p{K_0}{0}  - 2 \p{K_0}{2}\Big] \Ket{\p{n}{0}}.
		\end{align}
		Once again, it is evident that the corrections to the eigenstates depend only on $K_0$, highlighting the gauge dependence of the eigenstates.

		Our target is to determine the gauge-invariant quantity, namely the corrections to the Hamiltonian eigenvalues, which can be computed using Eq.~\eqref{K11}.
		 For order $q^{k - 1}$, the corresponding equation is
		\begin{equation}
			\sum_{j = 1}^{k} \Big(\p{K_1}{j - 1} - j \p{h_n}{j}\Big)\Ket{\p{n}{k - j}} = 0.	
		\end{equation}
		By sandwiching this equation with $\Bra{\p{n}{0}}$, we derive the recursion formula for the corrections to the eigenvalues, namely,
		\begin{align}
			\notag \p{h_n}{k} & = \frac{\dsb{\p{K_1}{k - 1}}{nn}}{k}\\
			& \quad + \sum_{j = 1}^{k - 1}\frac{\dsb{\p{K_1}{j - 1} \mathbb{B}_{k - j}}{nn} - \p{h_n}{j} \dsb{ \mathbb{B}_{k - j}}{nn}}{(k - j)! k},
\label{recurv}
		\end{align}
		where the double square brackets are defined as
		\begin{align}
			\dsb{\cdots}{mn} & \equiv \Bra{\p{m}{0}} \cdots \Ket{\p{n}{0}} = \bra{\p{m}{0}} \p{G}{0} \cdots \ket{\p{n}{0}}.
		\end{align}
		We list some results of the Hamiltonian eigenvalue corrections for reference,
		\begin{align}
			\notag \p{h_n}{1} & = \dsb{\p{K_1}{0}}{nn},\\[2mm]
			\notag \p{h_n}{2} & = \frac{1}{2}\Big(\dsb{\p{K_1}{1}}{nn} - i\dsb{\p{K_1}{0} \p{K_0}{0}}{nn}+ i \dsb{\p{K_0}{0}}{nn} \p{h_n}{1}\Big)\\[2mm]
			\notag \p{h_n}{3} & = \frac{1}{6}\Big(2\dsb{\p{K_1}{2}}{nn} - 2 i \dsb{\p{K_1}{1} \p{K_0}{0}}{nn} -i \dsb{\p{K_1}{0} \p{K_0}{1}}{nn} \\
			\notag &~~~~~ \quad - \dsb{\p{K_1}{0}\left(\p{K_0}{0}\right)^2 }{nn}+ i \dsb{\p{K_0}{1}}{nn} \p{h_n}{1}\\
			\label{energyc1}&~~~~~ \quad  + \dsb{\left(\p{K_0}{0}\right)^2 }{nn} \p{h_n}{1}+ 4 i \dsb{\p{K_0}{0}}{nn} \p{h_n}{2}\Big).
		\end{align}
		Generally speaking, we can compute the $K_0$ and $K_1$ operators to obtain the corresponding eigenvalue corrections order by order.

	\section{Special Case: Linear Perturbation}\label{secIV}

		In this section, we apply the aforementioned setup to the most commonly encountered situation, namely, a linear perturbation:
		\begin{equation}
			\label{linear} H = \p{H}{0} + q \p{H}{1}.
		\end{equation}

		By expanding Eqs.~\eqref{HK0} and \eqref{HK1} in powers of $q$, i.e.,
		\begin{align}
			& \bigg[\p{H}{0} + q \p{H}{1}, ~{\displaystyle\sum_{j = 0}^\infty} q^j \p{K_0}{j}\bigg] = i {\sum_{k = 0}^\infty} q^k \p{K_1}{k} - \p{H}{1},\\
			\notag~\\[-0.2cm]
			& \bigg[\p{H}{0} + q \p{H}{1}, ~{\displaystyle\sum_{j = 0}^\infty} q^j \p{K_1}{j} \bigg] = 0.
		\end{align}
		By matching the coefficients of each order of $q$, the above equations become
		\begin{align}
			& \left[\p{H}{0}, \p{K_0}{0}\right] = i \left(\p{K_1}{0} - \p{H}{1}\right) \label{KandHExpansion1}\\[1mm]
			& \left[\p{H}{0}, \p{K_1}{0} \right] = 0 \label{KandHExpansion2}\\[1mm]
			& \left[\p{H}{0}, \p{K_0}{\ell + 1}\right] + \left[\p{H}{1}, \p{K_0}{\ell}\right] = i \p{K_1}{\ell + 1} \label{KandHExpansion3}\\[1mm]
			&\left[\p{H}{0}, \p{K_1}{\ell + 1} \right] + \left[\p{H}{1}, \p{K_1}{\ell} \right] = 0, \label{KandHExpansion4}
		\end{align}
		for $\ell > 0$. Using Eqs.~\eqref{KandHExpansion1}-\eqref{KandHExpansion4} and the fact that
		\begin{align}
			 \dsb{AB}{nn} = \sum_m \dsb{A}{nm} \dsb{B}{mn},\label{Completeness}
		\end{align}
		 it is straightforward to show that the first three corrections to the eigenvalues shown in Eq.~\eqref{energyc1} become
		\begin{align}
			\notag \p{h_n}{1} & = \dsb{\p{K_1}{0}}{nn},\\[3mm]
			\label{energyc2} \p{h_n}{2} & = \frac{1}{2}\dsb{\p{K_1}{1}}{nn},\\[1mm]
			\notag \p{h_n}{3} & = \frac{1}{3}\dsb{\p{K_1}{2}}{nn} - \frac{1}{3}\sum_{m\neq n}\frac{\dsb{\p{K_1}{1}}{nm}\dsb{\p{K_1}{1}}{mn}}{\p{h_m}{1} - \p{h_n}{1}},
		\end{align}
		where the gauge-dependent $\p{K_0}{j}$'s either canceled with each other or were replaced by the gauge-independent $\p{K_1}{\ell}$'s.

		As a matter of fact, Eqs.~\eqref{KandHExpansion1}-\eqref{KandHExpansion4} also allow the perturbative corrections in Eq.~\eqref{energyc2} to be expressed in terms of $\p{H}{1}$. For instance,  Eq.~\eqref{KandHExpansion1} renders
		\begin{align}
			\begin{split}
				 \p{h_n}{1} & = \dsb{\p{K_1}{0}}{nn}\\
				 & = \dsb{\left(\p{H}{1} - i \left[\p{H}{0}, \p{K_0}{0}\right]\right)}{nn}\\
				 & = \dsb{\p{H}{1}}{nn} - i \left(\p{h_n}{0} - \p{h_n}{0}\right)\dsb{\p{K}{0}}{nn}\\
				 & = \dsb{\p{H}{1}}{nn},
			 \end{split}
		\end{align}
		where we have used $\p{H}{0} \Ket{\p{n}{0}} = \p{h_n}{0} \Ket{\p{n}{0}}$ and $\Bra{\p{n}{0}} \p{H}{0} = \p{h_n}{0} \Bra{\p{n}{0}}$ in the calculation. Analogously, straightforward calculations show that
		\begin{widetext}
\vspace{-3mm}
			\begin{align}
				\begin{split}
					\p{h_n}{1} & = \Bra{\p{n}{0}} \p{H}{1} \Ket{\p{n}{0}},\\[3mm]
					\p{h_n}{2} & = \sum_{m \neq n} \frac{\Bra{\p{n}{0}} \p{H}{1} \Ket{\p{m}{0}} \Bra{\p{m}{0}} \p{H}{1} \Ket{\p{n}{0}}}{\p{h_n}{0} - \p{h_m}{0}},\\[1mm]
					\p{h_n}{3} & = \sum_{\substack{m \neq n\\\ell \neq n}} \frac{\Bra{\p{n}{0}} \p{H}{1} \Ket{\p{m}{0}} \Bra{\p{m}{0}} \p{H}{1} \Ket{\p{\ell}{0}} \Bra{\p{\ell}{0}} \p{H}{1} \Ket{\p{n}{0}}}{\left(\p{h_n}{0} - \p{h_m}{0}\right) \left(\p{h_n}{0} - \p{h_\ell}{0}\right)} - \p{h_n}{1} \sum_{m \neq n} \frac{\Bra{\p{n}{0}} \p{H}{1} \Ket{\p{m}{0}} \Bra{\p{m}{0}} \p{H}{1} \Ket{\p{n}{0}}}{\left(\p{h_n}{0} - \p{h_m}{0}\right)^2}.
				\end{split}\label{GeneralizedPerturbation}
			\end{align}
		\end{widetext}
		Some important ingredients for deriving the equations above can be found in Appendix~\ref{App:Inggredients}.

		It is worth noting that if $\p{H}{0}$ is a Hermitian Hamiltonian, the above perturbation corrections can be reduced to the familiar results from the Rayleigh-Schrödinger perturbation theory by taking $\p{G}{0} = \mathbbm{1}$, so that $\Ket{\p{n}{0}} = \ket{\p{n}{0}}$ and $\Bra{\p{n}{0}} = \bra{\p{n}{0}} \p{G}{0} = \bra{\p{n}{0}}$.

		Therefore, the proposed perturbation scheme not only reproduces the well-known results in the Hermitian regime, but also extends its validity into the non-Hermitian regime. (Note that the proposed scheme fails when $\p{H}{0}$ is at an EP because Eq.~\eqref{Completeness} is not valid in this case.)

	\section{A Non-Trivial Example}\label{secV}
		Here we apply our formalism to a toy model,
		\begin{align}
			H = \p{H}{0} + q \p{H}{1} + q^2 \p{H}{2},
		\end{align}
		where $q$ is the expansion parameter,  and
		\begin{align}
			&\hspace{1.7cm} \p{H}{0} = h \left( \begin{array}{cc}
				0 & 1 \\
				1 & 0
				\end{array}\right), \\[2mm]
	\notag			& \p{H}{1}= \alpha_1 \left( \begin{array}{cc}
				i & 0 \\
				0 & - i
			\end{array}\right),~~~ \p{H}{2} = \alpha_2 \left( \begin{array}{cc}
				0 & 1 \\
				1 & 0
				\end{array}\right),
		\end{align}
		where $h$, $\alpha_1$, and $\alpha_2$ are fixed constants.

		Solving Eqs.~\eqref{HK0} and \eqref{HK1} iteratively with the expansion Eq.~\eqref{Expn}, we have
		\begin{align}
			\notag \p{K_0}{0} & = \left( \begin{array}{cc}
				 a_1 & - \dfrac{\alpha_1 }{2 h} + a_2 \\
				\dfrac{\alpha_1 }{2 h} + a_2& a_1
			\end{array}\right),\\[2mm]
			\notag K_1^{(0)} & = \left( \begin{array}{cc}
				 0 & 0 \\
				 0 & 0
			\end{array}\right),\\[2mm]
			\notag \p{K_0}{1} & = \left( \begin{array}{cc}
				 a_3 & a_4 \\
				 a_4 &   - 2 i a_2 \dfrac{\alpha_1}{h}+a_3
			\end{array}\right),\\[2mm]
			\label{exampleK} K_1^{(1)} & = - \frac{\alpha_1^2 - 2 \alpha_2 h}{h} \left( \begin{array}{cc}
				0 & 1\\
				1 & 0
			\end{array}\right),\\[2mm]
			\notag \p{K_0}{2} & = \begin{pmatrix}
				 a_5 & - \dfrac{\alpha_1^3 - 3 \alpha_1\alpha_2 h}{2h^3} + a_6 \\
				\dfrac{\alpha_1^3 - 3 \alpha_1\alpha_2 h}{2 h^3} + a_6 & - 2 i a_4 \dfrac{\alpha_1}{h} + a_5
			\end{pmatrix},\\[2mm]
			\notag \p{K_1}{2} & = - i \frac{ \alpha_1^3 - 2 h\alpha_1\alpha_2}{h^2} \left( \begin{array}{cc}
				 1 & 0 \\
				 0 & - 1
			\end{array}\right),
		\end{align}
		where the $a_i$'s are constants that correspond to the gauge degrees of freedom, which contribute only to the amplitudes of the eigenstates but not the eigenvalues. Direct calculation using Eq.~\eqref{exampleK}, the fact that $\p{H}{0}$ is Hermitian so that $\p{G}{0} = \mathbbm{1}$,  and the unperturbed eigenstates from $\p{H}{0}$, 
		\begin{equation}
			\label{examplestates}\Ket{\p{+}{0}} = \frac{1}{\sqrt{2}}\left( \begin{array}{c}
				1 \\
				1 
			\end{array}\right),~~~\Ket{\p{-}{0}} = \frac{1}{\sqrt{2}}\left( \begin{array}{c}
				1 \\
				- 1 
			\end{array}\right),
		\end{equation}
		yields
		\begin{equation}
			\p{h_\pm}{1} = 0,~~~\p{h_\pm}{2}= \mp q^2\left(\frac{\alpha_1}{2h} - \alpha_2\right), ~~~\p{h_\pm}{3} = 0.
		\end{equation}
		Notice that the $a_i$'s do not appear in the eigenvalue perturbation, as expected. As shown here, we provide a systematic method for computing the perturbations order by order.

	\section{Conclusions}\label{secVI}
		In this paper, we develop a perturbation scheme based on the geometric formalism~\cite{Mostafazadeh2004, Ju2019}. In addition to perturbing the states, we also perturb the matrices $K_0$ and $K_1$, which are related to their ascendant, the metric of the Hilbert space $G$. The perturbation in eigenstates can then be characterized by the recursion equations Eq.~\eqref{GNFstate}.

		As the recursion relations of eigenstates take a form similar to the well-known Girard-Newton formulas, and the Girard-Newton formulas can be related to  Bell polynomials, this observation inspires us to express each order of eigenstate correction in terms of dual non-commutative Bell polynomials. With these results, we further derive the recursion equation for eigenvalues Eq.~\eqref{recurv}. This result can be applied to any type of polynomial perturbation.  

As a consistency check, when considering the linear perturbation of a Hermitian Hamiltonian, Eq.~\eqref{linear}, the results in Eq.~\eqref{energyc1} and Eq.~\eqref{energyc2} are equivalent to those obtained from Rayleigh-Schrödinger perturbation theory. Nevertheless, for concreteness, a simple example with linear and quadratic perturbations is provided. It is straightforward to consider other polynomial perturbations as well.

In conclusion, the perturbation scheme presented in this paper can be applied to Hermitian systems, $\mathcal{PT}$-symmetric systems, pseudo-Hermitian systems, and even beyond.  However, the method generally does not work at exceptional points (EPs).  On one hand, the non-trivial metric of Hilbert space of non-Hermitian systems becomes singular at the EPs~\cite{Ju2024a}.  On the other hand,  we need to generalize our method for the case where the eigenvalues are degenerate and the eigenvectors coalesce.

Recent studies~\cite{Dembowski2004, Arkhipov2023, Lai2024, Beniwal2024} on the pertubation around EPs in non-Hermitian system primarily relied on the series expansions of exact eigenstates to derive perturbed results, rather than perturbing the system from unperturbed eigenstates.  To advance research in this area, the development of more general and systematic methods remains highly anticipated.

Lastly, since the resemblance to the Girard-Newton formulas has been incorporated as a novel element in this perturbation formalism,  it would be interesting to explore whether this new ingredient can provide additional input to the quantum mechanics bootstrap program~\cite{Han:2020bkb,Berenstein:2021loy,Li:2022prn,Khan:2024mhc}.

	\begin{acknowledgments}
		W.M.C. is partially supported by the National Science and Technology Council (NSTC) through Grant No. NSTC 113-2811-M-110-008. Both W.M.C. and C.Y.J. are partially supported by the NSTC through Grant No. NSTC 112-2112-M-110-013-MY3 and the National Center for Theoretical Sciences.
	\end{acknowledgments}

	\begin{appendix}
		\section{Girard-Newton Formulas and Bell Polynomials \label{App:GNFandBP}}
			As mentioned in the main text, Eq.~\eqref{GNFstate} takes a form similar to the so-called Girard-Newton formulas. For clarity,
			we review the Girard-Newton formulas in the appendix.
			Consider a monic polynomial $f(x)$ of degree $k$ with roots $x_1$, $x_2$,$\dots$, $x_k$, 
			\begin{equation}
				\label{fx} f(x) = \prod_{j = 1}^k(x - x_j) = \sum_{j = 1}^k s_{j - k} x^{j},~~\mathrm{with}~s_0 = 1 ,
			\end{equation}
			where, in the second equality,  the equation is rewritten using the  \textit{elementary symmetric polynomials},
			\begin{equation}
				\label{spoly}s_j(x_1,\dots,x_k) = ( - 1)^j \sum_{\resizebox{2cm}{!}{$1\leq m_1<\dots<m_j\leq k$}}x_{m_1}\cdots x_{m_j},
			\end{equation}
			The $s_i$'s form a basis for every symmetric polynomials in $x_1$, $x_2$,$\dots$, $x_k$. where as the \textit{power sums} 
			\begin{equation}
				p_i(x_1,\dots,x_k) = \sum_{j = 1}^k x^i_{j},
			\end{equation}
			form another basis. The Girard-Newton formulas gives a relationship between the two bases as
			\begin{equation}
				\label{GNF} s_k = - \frac{1}{k}\sum_{j = 1}^k p_{k} s_{k - j} ,~~~~ k \geq 1.
			\end{equation} 
			Actually, one can express $s_j$'s ($j\neq 0$) in terms of $p_j$'s and $s_0$. For the purpose, the general form of $s_j$ can be represented by using
			the so-called \textit{Bell polynomials} $B_j(p_1,p_2,\dots,p_j)$,
			\begin{equation}
				\label{GNSol}s_j = \frac{1}{j!}B_j( - p_1, - (1)!p_2,\dots, - (j - 1)! p_j)s_0.
			\end{equation}
			We refer the reader to ~\cite{Schimming96} and therein for the details of Bell polynomials and their non-commutative generalization mentioned here.
			By definition, the Bell polynomials satisfy the recursion equation (we omit the argument $p_1$,$\dots$,$p_k$ for simplicity),
			\begin{equation}
			\label{BDef} B_{k} = \sum_{j = 0}^k \left(k\atop j\right)B_{k - j}p_{j + 1} ~~\mathrm{with}~~B_0 = 1~~\mathrm{and}~~k\geq 0.
			\end{equation}
			For the generalizing to the case when $p_i$'s become non-commutative variables, it is good to decompose a Bell polynomial $B_k$ into the sum of \textit{homogeneous Bell polynomials} $B_{k,j}$,
			\begin{equation}
				\label{NCB} B_k = \sum_{j = 1}^k B_{k,j},
			\end{equation}
			~\\[-2mm]
			where for a collection of integers $m_1,\dots,m_k\geq 0$,
			\begin{align}
				& B_{k,j}(p_1,p_2,\dots,p_k) = k!\sum_{|m| = j\atop{||m|| = k}} \prod_{r = 1}^{k}\frac{1}{r!}\frac{(p_{m_r})^{m_r}}{m_r !},\\[2mm]
				\notag & ~~~~~~~~~~~~|m|\equiv m_1 + m_2 + \dots + m_k, \\
				\notag & ~~~~~~~~~||m||\equiv m_1 + 2m_2 + \dots + k m_k.
			\end{align}
			The definition of Bell polynomials, Eq.~\eqref{BDef} can be generalized to non-commutative Bell polynomials $\mathbb{B}_k$ by promoting $p_j$'s to some matrices $\mathbb P_j$'s, 
\vspace{-0.3mm}
			\begin{equation}
					\label{NCBDef} \mathbb{B}_{k} = \sum_{j = 0}^k \left(k\atop j\right)\mathbb{B}_{k - j}\mathbb{P}_{j + 1} ~~\mathrm{with}~~\mathbb{B}_0 = \mathbbm{1}~~\mathrm{and}~~k\geq 0.
			\end{equation}
			$\mathbb{B}_{k}$ is the \textit{dual non-commutative Bell polynomials} defined in ~\cite{Schimming96}. 
			The solution of $\mathbb{B}_{k}$ to the recursion equation is
\vspace{-0.5mm}
			\begin{equation}
				\mathbb{B}_{k} = \sum_{j = 1}^k \mathbb{B}_{k,j},
			\end{equation}
			where, 
			\begin{itemize}
				\item for $j = 1$,
				\vspace{1mm}
				\begin{equation}
				\hspace{-3.2cm}	\mathbb{B}_{k,1}(\mathbb{P}_1,\mathbb{P}_2\dots,\mathbb{P}_k) = \left(\mathbb{P}_1\right)^k,
				\end{equation}
				\item for $1<j<k$,
				\begin{align}
					& \mathbb{B}_{k,j}(\mathbb{P}_1,\mathbb{P}_2\dots,\mathbb{P}_k)\\
					\notag & = \sum_{m_2\dots m_j = 1}^k\left(k - 1\atop m_2\right)\left(m_2 - 1\atop m_3\right)\cdots \left(m_{j - 1}\atop m_j\right)\\
					\notag &\hspace{2cm}\times\mathbb{P}_{k - m_2} \mathbb{P}_{m_2 - m_3}\cdots \mathbb{P}_{m_{k - 1} - m_{k}} \mathbb{P}_{m_k},
				\end{align}
				\item for $j = k$,
				\begin{equation}
			\hspace{-3.5cm}		\mathbb{B}_{k,k}(\mathbb{P}_1,\mathbb{P}_2\dots,\mathbb{P}_k) = \mathbb{P}_k.
				\end{equation}
			\end{itemize}
The dual non-commutative Bell polynomials can be used to represent the solution of perturbed eigenstates in the main text.

		\section{Important Ingredients for Reinterpreting Eigenvalue Corrections in the Linear Perturbation Scenario \label{App:Inggredients}}

			Some useful identities for the second and third order term for the perturbed eigenvalues are provided in this section.  For $m\neq n$, we have
			\begin{align}
				 \dsb{\p{K_0}{0}}{nm} & = - i\frac{\dsb{\p{H}{1}}{nm}}{\p{h_n}{0} - \p{h_m}{0}},\\[2mm]
				 \dsb{\p{K_0}{1}}{nm} & = \frac{2i (\p{h_m}{1} - \p{h_n}{1})\dsb{ \p{H}{1}}{mn}}{(\p{h_m}{0} - \p{h_n}{0})^2}\\[2mm]
				\notag &  + i\sum_{k\neq n\atop{k\neq m}}\frac{(2 \p{h_k}{0} - \p{h_n}{0} - \p{h_m}{0})\dsb{\p{H}{1}}{mk}\dsb{\p{H}{1}}{kn}}{(\p{h_k}{0} - \p{h_n}{0})(\p{h_k}{0} - \p{h_m}{0})(\p{h_m}{0} - \p{h_n}{0})}\\
				\notag &  + \frac{\left(\dsb{\p{K_0}{0}}{mm} - \dsb{\p{K_0}{0}}{nn}\right)\dsb{ \p{H}{1}}{mn}}{\p{h_m}{0} - \p{h_n}{0}} ,\\[2mm]
				 \dsb{\p{K_1}{0}}{nm} & = 0,\\[2mm]
				\dsb{\p{K_1}{1}}{nm} & = 2\sum_{m\neq n}\frac{\Big|\dsb{\p{H}{1}}{nm}\Big|^2}{\p{h_n}{0} - \p{h_m}{0}}.
			\end{align}
			Most of the non-diagonal matrix elements are conveniently expressed in terms of the matrix elements of $\p{H}{1}$, although the expression still contains $\dsb{\p{K_0}{0}}{jj}$, which ultimately cancels out, ensuring the gauge invariance of the eigenvalues.

Another important identity in the derivation is used in the derivation
			\begin{align}
				\dsb{\p{K_1}{2}}{nn} & = \sum_{m\neq n}(\p{h_m}{1} - 4 \p{h_n}{1})\frac{\Big|\dsb{\p{H}{1}}{mn}\Big|^2}{(\p{h_m}{0} - \p{h_n}{0})^2}
				\\
				\notag &~~~~ 
				+ \sum_{\substack{m \neq n\\k \neq n}} \frac{3\dsb{\p{H}{1}}{nm}\dsb{\p{H}{1}}{mk}\dsb{\p{H}{1}}{kn}}{(\p{h_m}{0} - \p{h_n}{0})(\p{h_k}{0} - \p{h_n}{0})}.
			\end{align}
With these ingredients, Eq.~\eqref{GeneralizedPerturbation} can be derived easily.
~\\
~\\
~\\
~\\
~\\
~\\
~\\
~\\
~\\
~\\
~\\
~\\
~\\
~\\
	\end{appendix}

	\bibliography{References}

\end{document}